\documentclass[aps,prl,superscriptaddress,twocolumn,showpacs,longbibliography,amsmath,amssymb,nofootinbib]{revtex4-2}
\usepackage[utf8]{inputenc}
\usepackage[T1]{fontenc}
\usepackage{microtype}
\usepackage{graphicx}
\usepackage{mathbbol}
\usepackage{amssymb}
\usepackage{dsfont}
\usepackage{comment}
\usepackage[usenames,dvipsnames,svgnames,table]{xcolor}
\usepackage[colorlinks,linkcolor=blue,citecolor=blue,urlcolor=blue]{hyperref}
\usepackage{siunitx}

\renewcommand{\i}{\mathrm{i}}

\def\equationautorefname~#1\null{Eq.~(#1)\null}

\newcommand{\vect}[1]{{\mathbf #1}}

\begin{document}
	\title{Limit cycles as stationary states of an extended Harmonic Balance ansatz}
	\date{\today}
	\author{Javier del Pino}
	\affiliation{Institute for Theoretical Physics, ETH Zürich, 8093 Zürich, Switzerland}
	\author{Jan Ko\v{s}ata}
	\affiliation{Institute for Theoretical Physics, ETH Zürich, 8093 Zürich, Switzerland}
	\author{Oded Zilberberg}
	\affiliation{Department of Physics, University of Konstanz, 78464 Konstanz, Germany}
	
	\DeclareGraphicsExtensions{.pdf,.png,.jpg}
	
\begin{abstract}
    A limit cycle is a self-sustained periodic motion appearing in autonomous ordinary differential equations. As the period of the limit cycle is a-priori unknown, it is challenging to find them as stationary states of a rotating ansatz. Correspondingly, their study commonly relies on brute-force time-evolution or on circumstantial evidence such as instabilities of fixed points. Alas, such approaches are unable to account for the coexistence of multiple solutions, as they rely on specific initial conditions. Here, we develop a multifrequency rotating ansatz with which we find limit cycles as stationary states.  We demonstrate our approach and its performance in the simplest case of the Van der Pol oscillator. Moving beyond the simplest example, we show that our method can capture  the coexistence of all fixed-point attractors and limit cycles in a modified nonlinear Van der Pol oscillator. Our results facilitate the systematic mapping of out-of-equilibrium phase diagrams, with implications across all fields of natural science.
\end{abstract}

\maketitle	

Limit cycles (LCs) are periodic solutions of nonlinear ordinary differential equations (ODEs) that persist indefinitely in time. Commonly, the oscillation period is a-priori unknown, with oscillations reflecting the autonomous balance between the various forces acting on the system. Limit cycles are ubiquitous in all scientific fields, including physics, biology, and engineering~\cite{Strogatz2000}, with broad applications, such as developing optical frequency combs and sensors ~\cite{kim2020emergence, PhysRevApplied.4.024016,piergentili2021absolute}, understanding biological oscillators~\cite{may1972limit,jewett1998refinement,turchin2013complex}, designing control systems \cite{isidori2013nonlinear}, as well as realizing neuromorphic computing \cite{chalkiadakis2022dynamical}. Limit cycles have recently gained significant attention in the realm of out-of-equilibrium physics with a variety of realizations in light-matter ensembles~\cite{Seibold2019, CarlonZambon2020, Marconi2020, Keeling2010, Bhaseen2012b, Lee2013, Piazza2015, Chiacchio2019, Kessler2019,peters2021limit, Kongkhambut2022}, parametric oscillators~\cite{Bello2019a, CalvaneseStrinati2020, Inagaki2021, parametricBook},
electromechanical systems~\cite{Chen2016,Houri2019,Houri2021,Brenig2022}, superconducting circuits~\cite{khan2018frequency}, and optomechanics~\cite{Metzger2008,Lorch2014a,Bakemeier2015,NavarroUrrios2017,Rudner_2020,PhysRevLett.129.063605}.


The concept of LCs is defined differently in different fields, leading to ambiguity in the literature~\cite{jenkins2013self}. Most definitions require the LC to be an isolated, self-sustained periodic motion, where the system is stationary, where its Fourier transform amplitudes remain time-invariant and are asymptotically stable. 
While some definitions allow for more complex solutions such as quasiperiodic or chaotic motions, the key feature of a LC is the absence of a preferred time origin and phase locking to external forces. This ``resilience" against external perturbations or noise is crucial for the role LCs play in  synchronization~\cite{pikovsky2001universal}, and in instabilities in fluid dynamics~\cite{Thomas2004, Petrov2012}.

The study of LCs poses enduring challenges, due to their initially unknown oscillation period and the nonlinear nature of the problem. Various methods have been proposed for their study, including brute-force ODE integration~\cite{parker2012practical} and bifurcation theory of unstable fixed points~\cite{han2006bifurcation,guckenheimer2013nonlinear}. While useful, these methods crucially depend on the choice of initial conditions or are applicable only for specific types of LCs. As such, they often fail to  capture the coexistence of multiple LCs in the system. Moreover, to date, the upper bound for the number of LCs in a polynomial vector field remains unknown~\cite{hilbert1900mathematische}. 

The Harmonic Balance Method (HBM)~\cite{Genesio1992, Guskov_2007, Luo_2012, Krack_2019, ourPackage} stands as a powerful technique for solving stationary (static and periodic) solutions of ODEs. The method relies on an ansatz assuming that the solution takes the form of a multi-frequency Fourier series. Then, a nonlinear optimization algorithm is used to determine the coefficients of the series that satisfy the boundary conditions and equations of motion. The HBM connects to the Krylov-Bogoliubov averaging method for solving nonlinear ODEs with harmonic dependence~\cite{bogoliubov1961asymptotic, Rand_2005}, effectively describing various effects in nanomechanics~\cite{Leuch2016, PhysRevResearch.4.013149, Bachtold2022}, optics~\cite{lugiato1984ii}, optomechanics~\cite{Burgwal2020} and nuclear magnetic resonance~\cite{buishvili1979application}.

\begin{figure}[t]
	\includegraphics[width=\linewidth]{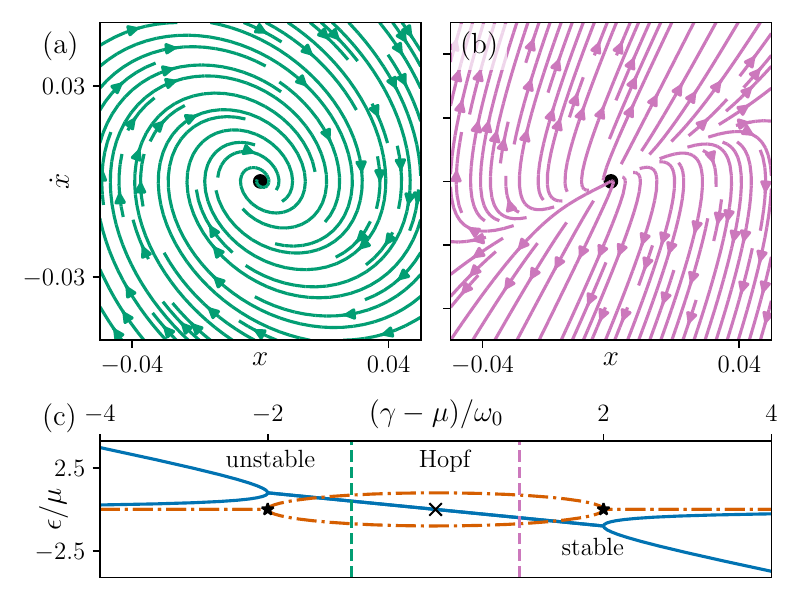}
	\caption{\textit{Gain versus loss and Hopf bifurcations.} Stream plots illustrating the behavior of a linear resonator with (a) positive loss, and (b) positive gain around the origin of parameter space (marked with black dot), cf.~Eq.~\eqref{eq:vdp} with $\alpha_{3,5}=0$ and $(\gamma -\mu)>0$ or $(\gamma -\mu)<0$, respectively. (c) A plot of the Jacobian eigenvalues around the origin $(0,0)$ as a function of linear damping for a Van der Pol oscillator [cf.~Eq.~\eqref{eq:vdp}]. Their imaginary part (orange, dashed-dotted lines) marks the oscillation frequency around the origin (excitation). Their real part (blue lines) denotes the inverse lifetime of the excitations (when negative) or their rate of expansion (when positive). From right to left, we observe transitions from overdamped to underdamped stable excitation (see the exceptional point at $\gamma-\mu=2\omega_0$, star marker), to unbound (unstable) spiral (Hopf bifurcation at 0, cross marker), and to an unbound non spiraling expansion (see the exceptional point at $\gamma-\mu=-2\omega_0$, star marker). Vertical dashed lines mark cases (a) and (b).} \label{fig:1}
\end{figure}

In this work, we introduce a variant of the HBM that can efficiently find LCs as stationary states. Specifically, we employ the HBM ansatz, and through gauge-fixing allow for a variational search of the LC frequencies using a powerful Homotopy continuation method~\cite{Sommese2005, Breiding_2018,timme2021numerical}. A substantial advantage of our approach is the ability to find the complete set of both LCs and fixed point stationary states under a generic ansatz, even when they coexist. We first illustrate the merit of our approach on the prototypical Van der Pol (VdP) oscillator, which is the simplest case study for LCs. We then generalize the model to include nonlinearities, where LCs and fixed point attractors can coexist. Crucially, our approach can be applied to efficiently explore the phase diagrams of nonlinear, coupled, driven-dissipative systems~\cite{CalvaneseStrinati2020, Alaeian2021,chiacchio2023non,dreon2022self}, including those commonly encountered in nanomechanics, cold atoms, and nonlinear photonics.

We consider a modified VdP oscillator that includes additional polynomial potential terms in the equation. The oscillator displacement $x(t)$ evolves according to Newton's equation $\ddot{x}+F(x,\dot{x})=0$ with
\begin{equation}\label{eq:vdp}
F(x,\dot{x})=\omega_0^2 x+\gamma\dot{x}-\mu(1-x^{2})\dot{x}+\alpha_3 x^{3}+\alpha_5 x^{5}\,,
\end{equation}
where the natural angular frequency is $\omega_0, \gamma$ controls the linear damping, $\mu>0$ is the standard VdP parameter that controls the linear gain channel alongside nonlinear damping, and $\alpha_{3,5}$ are cubic and quintic nonlinearities, respectively. 
The limit of  $\gamma,\alpha_{3,5}=0$ in Eq.~\eqref{eq:vdp} recovers the standard VdP equation, where the gain channel, $-\mu \dot{x}$, drives the oscillator to higher amplitudes, and the nonlinear damping, $\mu x^2\dot{x}$, saturates the motion into a limit cycle (cf.~the sign change of the overall term based on the value of $|x|$). In an opposite limit where only $\mu=0$, we have a nonlinear oscillator with quartic and sextic potential terms. This system can have multiple energy minima, which will be studied in the final part of the manuscript.

Our model has two distinct regimes related to the sign of $(\gamma -\mu)$. A positive sign signals a damped system that will relax into one of the potential minima depending on the initial boundary conditions, see Fig.~\ref{fig:1}(a). In the opposite case (negative sign), gain overcomes damping and the system is incoherently driven to higher amplitudes, see Fig.~\ref{fig:1}(b). Generally, for each point in configuration space, $\vect{x}_0=(x_0,\dot{x}_0)$, we can evaluate the Jacobian matrix $\mathcal{J}(\vect{x}_0)$ around that point to see where excitations flow to in the linear vicinity of $\vect{x}_0$. In particular, by diagonalizing the Jacobian at the origin, $\vect{x}_0=(0,0)$, we can discern between the two regimes above: the real part of the Jacobian eigenvalues   $\epsilon_{0}=\frac{1}{2}\left(-\left(\gamma-\mu\right)\pm\sqrt{(\gamma-\mu)^{2}-4\omega_0^2}\right)$ corresponds to whether the origin is a sink or a source (cf.~discussion on the sign of $\gamma-\mu$ above); the imaginary part parametrizes the oscillation frequency for small excitations near the origin, see Fig.~\ref{fig:1}(c). These small excitations are stable and overdamped when  $\gamma-\mu>2\omega_0$, and stable and underdamped when $0<\gamma-\mu<2\omega_0$. The under- to over-damped transition is marked by an exceptional point~\cite{Fernandez2018,khedri2022fate}. The source-to-sink transition occurs via a Hopf bifurcation, where the excitations lose stability while maintaining their oscillation frequency.  At higher gain, no notion of unstable excitations remains as the oscillation frequency is removed.

\begin{figure}
\includegraphics[width=\linewidth]{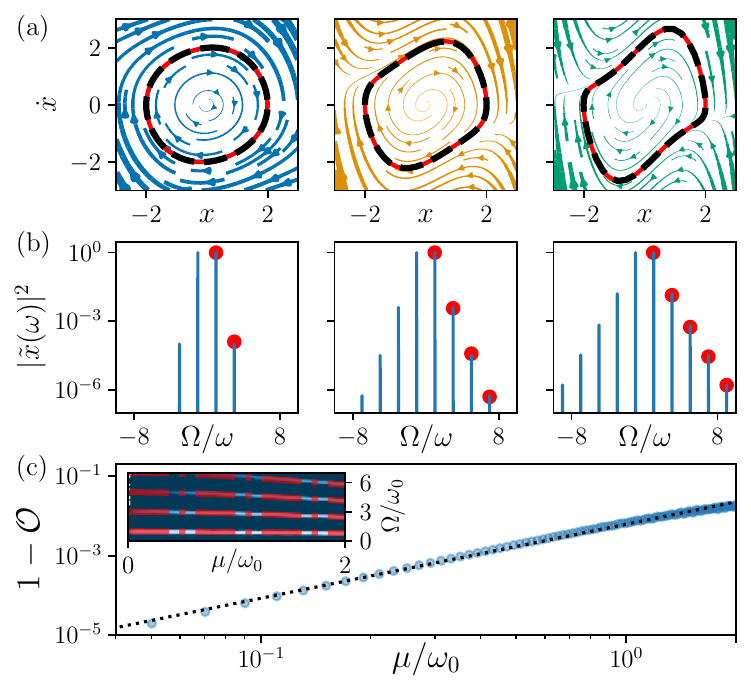}
	\caption{\textit{The van der Pol limit cycle and the harmonic balance method.} (a) Stream plots depicting the dynamics in the VdP oscillator [cf.~Eq.~\eqref{eq:vdp} with $\alpha_{3,5},\gamma=0$] for $\mu=0.1\omega_0$, $\mu=0.5\omega_0$, and $\mu=\omega_0$, from left-to-right, respectively. The solid (red) curve marks the LC obtained from direct time propagation, while the dashed (black) curve shows the result from the extended HBM with $M=5$ harmonics [cf.~Eqs.~\eqref{eq:LC_HBM}]. (b) Fourier transforms of the LC time evolutions in (a) unveil the emergence of a frequency comb. Red dots show normalized squared amplitudes at odd frequencies $(2k+1)\omega$ $k\in\mathbb{N}$, captured by the e-HBM. We use windowed Fourier transforms with a Hamming function to mitigate spectral leakage caused by finite integration times. (c) Overlap between e-HBM and exact time evolution result as a function of $\mu$ for $M=5$ [cf.~Eq.~\eqref{eq:overlap}]. Dots show a linear fit. (inset) Same as (b) as a function of $\mu$ marking the change in the comb frequencies. e-HBM frequencies are shown as dashed lines, overlaid on the positive part of the Fourier spectrum. 
   } \label{fig:2}
\end{figure}

The local (sink-to-source) instability around the origin does not provide sufficient information regarding whether a LC forms once the Hopf bifurcation occurs. Indeed, without any nonlinearity in the model, the excitations will expand under the gain to infinity. However, in the standard VdP oscillator ($\alpha_{3,5},\gamma=0$), the nonlinear damping prevents the system from spiraling to $x,\dot{x}\rightarrow\infty$, by stabilizing a LC that encircles the origin at a distance, see Figs.~\ref{fig:2}(a). Interestingly, the geometry of the LC is determined by the value of $\mu$. For smaller values, the motion exhibits a circular orbit in configuration space, which gradually morphs into a more complex pattern as $\mu$ increases. The periodic character of the LC makes it suitable for a Fourier decomposition, see Figs.~\ref{fig:2}(b). We observe that the oscillation frequency is \textit{not} simply that of the bare resonator ($\omega_0$). Specifically, the deviation from a single harmonic behavior (a circular motion in configuration space) is captured by the Fourier amplitudes $|\tilde{x}(\omega)|^2$, which reveal the formation of a frequency comb with harmonics $m\omega$. Note that the LC, in our case, has a half-wave symmetry in the time domain, such that only odd harmonics contribute.

Due to the intricate motion of LCs, with unknown frequencies, it is challenging to describe them as stationary states of frequency expansion, i.e., a rotating ansatz. Instead, LCs are commonly obtained by brute-force numerical integration, or by seeding them out of a Hopf bifurcation, cf.~Fig.~\ref{fig:1}. However, both methods have their limitations. The former may not capture the LC if the initial condition is not within its basin of attraction. The latter approach is not universal in capturing LCs: (i) a parameter deformation of the system that leads to a sink-to-source transition –– a necessary condition for a Hopf bifurcation –– is required; in the standard VdP with $\mu>0$, this transition is omitted. In addition, (ii) LCs can arise from various types of bifurcations, and not just Hopf, e.g., orbits connecting saddle equilibria, and even from infinity~\cite{Rand_2005, Guckenheimer_2013}. Consequently, in the following, we devise a variant of HBM~\cite{Krack_2019} that captures LCs as stationary states of a \textit{varying} rotating frame, i.e., at a time-evolving frequency to be discovered. 

In the standard HBM ansatz, the motion
\begin{equation}\label{eq:ansatz}
    x(t)= \sum_{l=1}^M u_l(t)\cos(\omega_l t) + v_l(t)\sin(\omega_l t)\,,
\end{equation}
is split into $M$ fast harmonics with frequencies  $\boldsymbol{\omega}=(\omega_1,\ldots,\omega_M)$ that have slowly evolving amplitudes $\vect{u}=(u_1(t),\ldots,u_M(t))$ and $\vect{v}=(v_1(t),\ldots,v_M(t))$. The ansatz~\eqref{eq:ansatz} approximates the time evolution with its leading frequency components. Finding the amplitudes $\vect{u}$ requires inserting it into Eq.~\eqref{eq:vdp}, which for $\alpha_{3,5}=0$ yields
\begin{widetext}
\begin{align}\label{eq:explicit_harm_subs}
    \sum_l\cos\left(\omega_{l}t\right)\left[\ddot{u}_{l}+2\omega_{l}\dot{v}_{l}-\mu\dot{u}_{l}+\left(\omega_{0}^{2}-\omega_{l}^{2}\right)u_{l}-\mu\omega_{l}v_{l}\right]+
    \sum_l\sin\left(\omega_{l}t\right)\left[\ddot{v}_{l}-2\omega_{l}\dot{u}_{l}-\mu\dot{v}_{l}+\mu\omega_{l}u_{l}+\left(\omega_{0}^{2}-\omega_{l}^{2}\right)v_{l}\right]=\\
    \sum_{m,n,l}\mu\left[u_{m}\cos\left(\omega_{m}t\right)+v_{m}\sin\left(\omega_{m}t\right)\right]
\left[u_{n}\cos\left(\omega_{n}t\right)+v_{n}\sin\left(\omega_{n}t\right)\right]\cdot
\hspace{2mm}\left[\sin\left(\omega_{l}t\right)\left(\omega_{l}u_{l}-\dot{v}_{l}\right)-\cos\left(\omega_{l}t\right)\left(\dot{u}_{l}+\omega_{l}v_{l}\right)\right]\,,\nonumber
\end{align}
\end{widetext}
We search for stationary motion implying fixed harmonic amplitudes, i.e., $\ddot{\mathbf{u}}=\ddot{\mathbf{v}}=\dot{\mathbf{u}}=\dot{\mathbf{v}}=0$. Furthermore, the premise of the HBM is to ``balance'' the harmonics at both sides of Eq.~\eqref{eq:explicit_harm_subs}, i.e., the prefactors of each harmonic satisfy the equation independently. Note that nonlinear terms in Eq.~\eqref{eq:vdp} lead to frequency mixing among different harmonics through 'multi-photon' resonance conditions, $\omega_l=\omega_l\pm\omega_m\pm\omega_n$. These conditions lead to third-degree nonlinear coupling among the prefactors of various harmonics.  As a result, we obtain from Eq.~\eqref{eq:vdp} a system of $2M$ coupled polynomial equations  $\mathcal{F}(\vect{u},\vect{v})$. When the stationary state frequencies $\omega_l$ are known, as in stationary motion fixed by an external driving, the HBM simplifies the problem to finding the roots $\mathcal{F}(\vect{u},\vect{v})=0$~\cite{supmat}.

\begin{figure*}[t]
	\includegraphics[width=\linewidth]{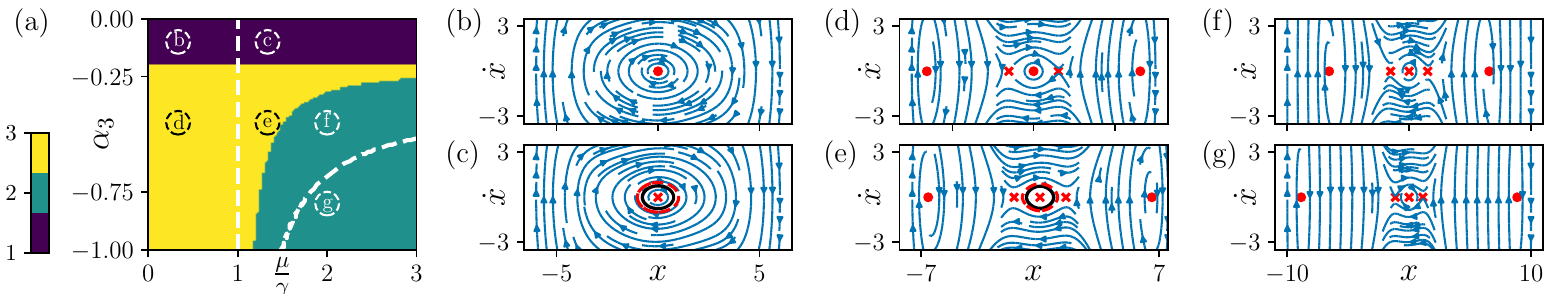}
	\caption{\textit{Coexistence of limit cycles and fixed points.} 
    (a) Phase diagram of the model~\eqref{eq:vdp} in the $\alpha_3,\mu$ parameter space, showing the count of stable stationary states, including LCs and fixed points. Dashed lines mark additional boundaries where the number of unstable stationary states changes. Panels (b)-(g) illustrate distinct behaviors in each region through stream plots [cf.~Fig.~\ref{fig:1}]]. Stable (unstable) fixed points derived from the HBM are depicted by dots (crosses). The LC found by the e-HBM is shown by a red dashed line. Regions (b) and (d) exist below the Hopf bifurcation, i.e., $\mu<\gamma$, and only sustain fixed points.  When $\mu>\gamma$, regions (c) and (e) host a LC around the origin, that coexists with stable fixed points in (e). The transition between (b) and (d) [respectively (c) and (e)] is associated with the appearance of two additional fixed points. The LC destabilizes in (f) and vanishes in (g) by merging with the unstable fixed points, and leaves behind solely fixed points. 
    In all panels, $\omega_0=1$, $\mu=0.1$, $\gamma=0.03$,  $\alpha_5-0.01$ and $M=2$.} \label{fig:3}
\end{figure*}

Limit cycles evolve with unknown self-oscillation frequencies. We, therefore, modify the standard HBM approach and introduce a slow time variation to the frequencies, $\omega_l\mapsto\omega_l(t)$, on top of the slowly evolving amplitudes $u_l(t),v_l(t)$ in the ansatz~\eqref{eq:ansatz}. In our extended HBM (e-HBM) approach, the variational prefactors evolve similarly to Eqs.~\eqref{eq:explicit_harm_subs}, with additional terms arising from time derivatives of $\omega_l(t)$. Now, fixing the harmonic amplitudes in the generalized stationary state, where $\ddot{\vect{u}}=\dot{\vect{u}}=\ddot{\vect{v}}=\dot{\vect{v}}=\ddot{\boldsymbol{\omega}}=\dot{\boldsymbol{\omega}}=0$, results in a system of $2M$ coupled polynomials 
$\tilde{\mathcal{F}}(\vect{u},\vect{v},\boldsymbol{\omega})=0$ in the $3M$ variables $\vect{u},\vect{v},\boldsymbol{\omega}$. The harmonic balance equations are thus insufficient to determine unequivocally the LCs. 
Fortunately, we can deduce the missing $M$ conditions from the spontaneous, self-oscillating nature of the LC: With no external drive to lock onto, the LC's phases are free from constraints. This freedom leaves the complex harmonic amplitude vector $\vect{a}=(\vect{u}+i\vect{v})/2$ with a free $U(1)$ gauge symmetry under global transformations $a_i\mapsto e^{i\varphi_i}a_i$, where $\varphi_i\in(0,2\pi)$~\cite{supmat}. As the stationary solution of the LC does not rely on the gauge, we can  fix the $M$ gauges by redefining the time origin in Eq.~\eqref{eq:ansatz}. One possible gauge choice yields, e.g., the following system of $3M$ equations
\begin{equation}\label{eq:LC_HBM}
\{\mathcal{F}(\vect{u},\vect{v},\boldsymbol{\omega})=0,\vect{u}=0\}\,.
\end{equation}
%
Note that our treatment simplifies when introducing relationships within $\boldsymbol{\omega}$, e.g. commensurate frequencies. For example, the LC in Eq.~\eqref{fig:2}(b) is described by a frequency comb with a single unknown frequency $\omega$, i.e., $\omega_l=l\omega$ with $l\in\mathbb{N}$.   This requires only fixing the gauge of a single harmonic to attain a solution, e.g., by setting $u_1=0$.

Our e-HBM embeds LCs as fixed points in a high-dimensional phase space, rotating with variational frequencies $\omega_l$. The complexity of solving the problem at this stage is delineated to the proliferation of roots in the system, which complicates the use of straightforward root-finding (e.g., Newton-Raphson) methods. Inspired by a similar challenge encountered in the standard HBM for driven motion~\cite{Krack_2019}, 
we solve Eqs.~\eqref{eq:LC_HBM} using Homotopy Continuation~\cite{Sommese2005, Breiding_2018,ourPackage,borovik2023khovanskii}. Homotopy Continuation combines root finding with continuous deformation of an exactly solvable algebraic system into the target coupled polynomials~\eqref{eq:LC_HBM} to find all the roots in a single run~\cite{Sommese2005, Breiding_2018}. Applying our methodology to the VdP oscillator ($\alpha_{3,5}=\gamma=0$), we find a single real solution that corresponds to the amplitudes $\vect{u},\vect{v}$ and frequencies $\boldsymbol{\omega}$ that well estimate the LC [Figs.~\ref{fig:2}(a)]. This same agreement manifests more clearly in the frequency domain [see Fig.~\ref{fig:2}(b)], and persists as a function of $\mu$ for various LC frequencies, see inset of~\ref{fig:2}(c).  To further benchmark the quality of the estimated solution, we compare the estimated e-HBM stationary amplitudes, and those obtained from the Fourier transform of the exact time-evolution. To this end, we evaluate the overlap 
\begin{align}
\mathcal{O}=\vect{a}_{\text{HBM}}^{*}\cdot \vect{a}_{\text{FT}}\,,
\label{eq:overlap}
\end{align}
between the normalized harmonic amplitudes $\vect{a}_{\text{HBM}}=(\vect{u}+i\vect{v})/\sqrt{\vect{u}^2+\vect{v}^2}$, and $\vect{a}_{\text{FT}}$, respectively. The overlap decreases with increasing $\mu/\omega_0$, approximately as $\sim(\mu/\omega_0)^2$ [cf.~Fig.~\ref{fig:2}(c)] reaching few-percent deviations for $\mu=2\omega_0$. The relationship between the overlap and the number of harmonics, $M$, is not strictly monotonic due to the specific orders retained in the HBM truncation~\cite{supmat}.

 The combination of HBM and Homotopy Continuation captures all solutions simultaneously, unlike direct time evolution, which only explores a single solution from a specific initial condition (a basin of attraction). While the system with $\alpha_{3,5}=0$ sustains a unique limit cycle, as guaranteed by Lienard's theorem~\cite{lienard1928etude}, when $\alpha_{3,5}\neq 0$ a coexistence of different stationary states manifests, see Fig.~\ref{fig:3}(a). Such coexistence arises from the conservative potential that hosts multiple energy minima, whereas around the origin the LC physics manifests. To describe this coexistence using e-HBM, we apply a two-step procedure: (i) we first solve for stationary fixed-points using a zero-frequency ansatz $x(t)=a_0(t)$, and then (ii) solve for the LCs using the e-HBM ansatz~\eqref{eq:ansatz}. Here, convergence is achieved by incorporating harmonics at frequencies $0$ and $(2l+1)\omega$ with $l\in\mathbb{N}$. Note also that the solution for $a_0$ recovers also the trivial fixed point in the VdP oscillator (the origin). The two steps are needed because the e-HBM ansatz leads to infinitely-degenerate solutions arising around fixed points, as it corresponds to zero amplitude solutions for arbitrary $\boldsymbol{\omega}\neq 0$. Importantly, by combining the results from the two steps, we fully estimate the expected outcome. Crucially, we can now use this combined e-HBM to capture the stationary solutions of the system (including LCs) for different parameters, e.g., as a function of the cubic potential and nonlinear gain, and obtain the phase diagram of the system, see Fig.~\ref{fig:3}(a) highlighting labeled distinct regions. Remarkably, a stable LC can coexist with fixed point attractors at region (e). 
 For the purpose of this work, we focus on the parameter regime where LCs coexist with fixed points, and do not study the potential mixing between the LC and attractors in the system. This coexistence exemplifies the loose connection between Hopf bifurcations and LCs, as no bifurcation appears, see e.g. transition between regions (e) to (f). 

Our method facilitates the  investigation of LCs in a breadth of fields and is readily available as part of HarmonicBalance.jl~\cite{ourPackage}. Instead of relying on numerical evolution and experimentation, it finds the LCs as fixed points of an extended rotating ansatz. As such, our treatment can be readily extended to non-autonomous systems, particularly those subjected to periodic forcing, where dynamics can undergo spontaneous shifts from periodic to quasi-periodic behavior \cite{Bello2019a}. By employing this approach, we expect to uncover the coexistence of solutions with stable phases, and also identify the presence of multiple limit cycles within the same system~\cite{kuznetsov2013visualization, PhysRevX.12.031003}. Crucially, our approach is well-suited for the analysis of many-body phases in mean-field problems~\cite{Aldana2013, Chitra2015,2021PhRvX..11d1046F}, and of the  conditions for synchronization in classical and quantum oscillator arrays~\cite{wachtler2023topological,moreno2023synchronized,lee2013quantum,walter2014quantum,walter2015quantum, PhysRevLett.113.154101, PhysRevLett.121.063601, PhysRevLett.120.163601, PhysRevLett.121.053601,BenArosh2021}. Notably, in all these fields, the ability to describe the LCs as a fixed point provides the framework for studying stochastic dynamics and activation between LCs. Last, while the e-HBM serves as a promising approach, its effectiveness is ultimately constrained by the solution of $2M$ coupled polynomial equations~\cite{Cox_2013}. Progress towards efficiently solving the equations~\eqref{eq:LC_HBM} will allow us to solve complex systems with more degrees of freedom~\cite{breiding2022algebraic}. 

\textit{Note added:} While preparing our manuscript, we encountered a manuscript that  proposes a related approach for LCs in aerodynamics~\cite{Petrov2012}, where a related HBM approach is used followed by a single solution continuation.


We thank A. Eichler, A. Nunnenkamp, and C.~W. Wächtler for fruitful discussions.  J.d.P. was supported by the ETH Fellowship program (grant no. 20-2 FEL-66). J.K. was supported by the Swiss National Science Foundation through grant CRSII5 $177198/1$. O.Z.  acknowledges funding through SNSF grant CRSII5\_206008/1, and the Deutsche Forschungsgemeinschaft (DFG) through project number 449653034, and via SFB1432.

\bibliography{refs}	
	
\clearpage
\pagebreak
\renewcommand{\theequation}{S\arabic{equation}}
\renewcommand{\thefigure}{S\arabic{figure}}
\setcounter{equation}{0}
\setcounter{figure}{0}
\setcounter{table}{0}
\setcounter{page}{1}

\onecolumngrid
\begin{center}
	\textbf{\large Supplemental Material: Limit cycles as stationary states of an extended Harmonic Balance ansatz}
\end{center}
\setcounter{secnumdepth}{2} 
\twocolumngrid

\section{Extended Harmonic Balance method}
The Harmonic Balance Method (HBM) is commonly applied to handle harmonically driven nonlinear systems. Such systems are described by general nonlinear system of $N$ second-order ordinary differential equations (ODEs) of the form
\begin{equation}\label{eq:generic_system}
	\dot{\textbf{x}}(t) = G(\textbf{x}(t), t),
\end{equation}
where $\textbf{x}(t)=(x_1(t),x_2(t),\cdots,x_N(t))^T$, and $x_i(t)$ are real variables with $i = 1, 2, ..., N$ and time $t$ as the independent variable. The function $G(\textbf{x}(t), t)$ is assumed to depend harmonically on $\textbf{x}(t)$ and its time derivatives\footnote{Note that in a linear system, $\mathbf{G}(\mathbf{x}, t) = \ddot{\mathbf{x}} + \mathbf{M}(t)\mathbf{x} + \mathbf{b}(t)$, where $\mathbf{M}(t)$ contains spring constants and linear couplings, and $\mathbf{b}(t)$ represents external forces. Diagonalizing $\mathbf{M}(t)$ yields the normal modes of the system.}. The system~\eqref{eq:generic_system} eventually reaches a stationary state, typically exhibiting an oscillatory response that remains constant over time. The HBM starts with the premise of representing such a solution with an expansion of the form [see main text Eq.~(2)]
\begin{align}\label{eq:HBM_ansatz}
	x_i(t)=\sum_{l=1}^{M_i} u_{l,i}(t)\cos(\omega_{l,i}t)+v_{l,i}(t)\sin(\omega_{l,i}t)\,,
\end{align}
where we introduced $M_i$ harmonics oscillating at frequencies $\omega_{l,i}$ with slowly-varying envelopes $u_{l,i}$ and $v_{l,i}$, in order to describe the evolution of each variable $x_i(t)$. An approach to finding such amplitudes is to perform numerical experiments, where ODE solvers propagate the evolution for a long time and Fourier analysis of the solutions is performed. Once a given set of solutions has been found for a parameter set, additional solutions can be obtained by continuation techniques~\cite{Krack_2019}. This methodology, however, can be challenging to apply, since nonlinear systems can have multiple stationary states, resulting in different responses depending solely on the initial conditions. Obtaining a complete stationary-state diagram requires time-consuming sampling of initial conditions, which may not guarantee the discovery of all branches of stationary solutions.

To simplify the process, an alternative approach involves using Eq.~\eqref{eq:HBM_ansatz} as an ansatz of the ODE system. Replacement of the expansion~\eqref{eq:HBM_ansatz} into ~\eqref{eq:generic_system} leads to harmonic oscillating terms at both sides of the equation. The stationary state problem can be transformed into an algebraic one, within the truncated ansatz~\eqref{eq:HBM_ansatz}, by `balancing' harmonics at both sides of Eq.~\eqref{eq:generic_system}. This procedure results in equations equivalent to those found by Krylov-Bogoliubov averaging~\cite{Rand_2005} (see main text and Ref.~\cite{ourPackage} for further details):
\begin{equation}
	\mathcal{F}(\textbf{u},\textbf{v})=0,
\end{equation}
where $\textbf{u}=(u_{1,1},\cdots,u_{M_N,N})$, $\textbf{v}=(v_{1,1}\cdots,v_{M_N,N})$.

In typical scenarios, an externally driven system settles into stationary motion states that oscillate at frequencies that are equal or commensurate with the driving frequencies. Therefore, the frequencies $\omega_{l,i}$ can be assumed a priori (e.g., they are prescribed by the driving frequencies), while the variables $\textbf{u},\textbf{v}$ remain unknown. In limit cycle (LC) problems, however, this is no longer the case, since the complex interplay of driving, dissipation, and nonlinearity determines the oscillation frequencies. In order to solve such frequencies, we resort to the concept of gauge freedom in LCs. 
Gauge freedom fundamentally stems from the spontaneous emergence of LCs, which causes a striking change in the time-translation symmetry of the problem: a time-dependent solution can arise even in a time-independent system. We illustrate how this concept allows to extract limit cycles as the non-redundant stationary solutions from an HBM ansatz, upgrading the HBM into the extended HBM (e-HBM).

In the main text, the concept of gauge freedom enables the determination of the system of polynomial equations for the stationary state by identifying a specific, non-redundant solution. Here, we delve into the emergence of gauge freedom and discuss in the case of $N=1$ to keep the discussion simple. We deduce the notion of gauge freedom from the LC symmetry under time translations with the ODE's period $T_s$, namely the operation $\mathcal{T}(T_s)x(t)=x(t+T_s)$, or equivalently, $\mathcal{T}(T_s)a_m=a_me^{im\omega_m T_s}$. 
A periodic nonlinear system is invariant under the \textit{discrete} operation $\mathcal{T}(T_s)$, so it is a time-independent (autonomous) system with a continuous, arbitrary $T_s$. However, their long-term solutions can \textit{spontaneously} break these symmetries. For instance sub-harmonic oscillations with commensurate periods $T_sq/p$ ($p,q\in\mathbb{N}$) break $\mathcal{T}(T_s)$ symmetry. This allows all solutions returning to themselves after $q$ iterations of the original translation $\mathcal{T}(T_s)$, which are related via discrete rotations  $a_m\rightarrow e^{i2\pi j/q} a_m$ for $j\in\{0,1,\cdots,q-1\}$. However, a \textit{continuous} time-translation-symmetry broken LC has an incommensurate period $T_s r$ with $r$ irrational; mathematically, LC orbits only return to themselves after infinite actions of $\mathcal{T}(T_s)$ with fixed $T_s$. This property frees oscillation phases in LCs from constraints and yields a continuous set of solutions related by full $U(1)$ symmetry $a_i\mapsto e^{i\varphi_i}a_i$, where $\varphi_i\in(0,2\pi)$.

\begin{widetext}
	\section{Convergence of the e-HBM for the Van der Pol oscillator}
	
	Here, we provide details on the application of the e-HBM to the Van der Pol (VdP) oscillator in the main text. For the sake of simpler algebraic manipulation, we adopt a complex notation where $\textbf{a}=\textbf{u}+i\textbf{v}$, thus writing the HBM ansatz as
	\begin{equation}\label{eq:vdp_comm}
		x(t)=\sum_{m=1}^{M}a_{m}(t)e^{-im\omega t}+\mathrm{H.c.=}\sum_{k=0}^{\frac{M-1}{2}}a_{k}(t)e^{-i(2k+1)\omega t}+\mathrm{H.c.}\,.
	\end{equation}
	The first step of the HBM involves inserting Eq.~\eqref{eq:vdp_comm} into the VdP equation, namely
	\begin{equation}
		\ddot{x} - \mu(1-x^2)\dot{x} = 0\,,
	\end{equation}
	as we illustrated in Eq.~(3) of the main text. Next, we match oscillatory terms with a specific frequency $\omega_k$ and isolate the corresponding frequency terms. Such frequency matching procedure filters each oscillating contribution in an 
	equivalent way as averaging the equation over the period $2\pi/\omega_k$, such that $\omega_k=k\omega$.
	
	The nonlinearity in the VdP equation leads to couplings between different harmonic amplitudes. We focus on the impact of this nonlinear damping in the VdP oscillator. The corresponding term in the Harmonic Balance equation at a frequency $\omega_{k}$ follows from the average integral:
	\begin{equation}
		\langle x^{2}\dot{x}\rangle^{c}=-i\frac{\omega_{k}}{2\pi}\int_{-2\pi/\omega_{k}}^{2\pi/\omega_{k}}\mathrm{d}t\sum_{m,n,p}\omega_{m}\left[a_{m}e^{-i\omega_{m}t}-a_{m}^{*}e^{i\omega_{m}t}\right]\left[a_{n}e^{-i\omega_{n}t}+a_{n}^{*}e^{i\omega_{n}t}\right]\left[a_{p}e^{-i\omega_{p}t}+a_{p}e^{i\omega_{p}t}\right]e^{-i\omega_{k}t}\,.
	\end{equation}
	with $a_m(t)$ assumed constant for the purpose of the time integration [see~\cite{ourPackage}].
	
	The equations for the harmonic amplitudes $u_k$ and $v_k$ can be derived by separating the real and imaginary terms from the aforementioned expression. The key point is that, for each $k$, all terms with $m,n\leq M$ are present, and explicitly, the surviving terms are
	\begin{align}\label{eq:expansion_average}
		\langle x^{2}\dot{x}\rangle_{k}^{c}=i\sum_{m,n,p}\omega_{m}\left[a_{m}a_{n}\left[a_{p}\delta_{m+n+p+k}+a_{p}^{*}\delta_{m+n-p+k}\right]-a_{m}^{*}a_{n}\left[a_{p}\delta_{-m+n+p+k}+a_{p}^{*}\delta_{-m+n-p+k}\right]\right]+\nonumber\\	
		i\sum_{m,n,p}\omega_{m}\left[a_{m}a_{n}^{*}\left[a_{p}\delta_{m-n+p+k}+a_{p}^{*}\delta_{m-n-p+k}\right]-a_{m}^{*}a_{n}^{*}\left[a_{p}\delta_{-m-n+p+k}+a_{p}^{*}\delta_{-m-n-p+k}\right]\right]\,,
	\end{align}
	where $\delta_{m-n} = \frac{1}{2\pi} \int_{-\pi}^{\pi} e^{i(m-n)\theta} \, d\theta$ stands for the Kronecker symbol. The preceding step assumes the commensurability of frequencies, which has been previously assumed in the Van der Pol ansatz Eq.~\eqref{eq:vdp_comm}. The ansatz insertion, expansion, and frequency balance are facilitated in the main text by the symbolic layer of the HarmonicBalance.jl package~\cite{ourPackage}.
	
	Equation~\eqref{eq:expansion_average} reveals explicit nonlinear coupling between different harmonics, weighted by `selection rules' arising from the Kronecker delta factors.
	Under the assumption of convergence of the expansion Eq.~\eqref{eq:vdp_comm}, we expect some of the harmonics to play a more predominant role than others: in practice, we observe the harmonic amplitudes to monotonically decay, cf.~Fig.~2(b) in main text. Therefore, to assess convergence, it is instrumental to analyze how higher-frequency harmonics impact low-frequency ones. For this analysis, it is important to also notice that the LC in the VdP oscillator exhibits a series of step functions in the time domain that oscillate with half-wave symmetry with period $T$, 
	($x(t)=-x(t\pm T/2)$), i.e., only odd harmonics contribute to the ansatz. The sums $\sum_{m,n,p}$ must be thus restricted to odd harmonics only, so we replace $m\rightarrow2m+1,n\rightarrow2n+1,p\rightarrow2p+1,k\rightarrow2k+1$ in the harmonic frequencies and set the origin in the indices to zero: $m,n,p,k\geq0$.  
	
	Let as analyze the terms that arise from a given truncation $M$. Keeping only the first harmonic in the ansatz, i.e., setting $M=1$ and collecting terms with $m=n=p=k=0$, the only non-vanishing contributions in \autoref{eq:expansion_average} are
	\begin{align}
		-a_{m}^{*}a_{n}^{*}a_{p}\delta_{2(k-m-n+p)}=-a_{0}^{*}|a_{0}|^{2},&&
		-a_{m}^{*}a_{n}a_{p}^{*}\delta_{2(k-m+n-p)}=-a_{0}^{*}|a_{0}|^{2},&&	
		a_{m}a_{n}^{*}a_{p}^{*}\delta_{2(k+m-n-p)}=a_{0}^{*}|a_{0}|^{2}\,.	
	\end{align}
	At this approximation level, equivalent to the lowest-order averaging method [see~\cite{Rand_2005} for further details], the nonlinear dissipation in the VdP oscillator appears as an amplitude-dependent dispersive shift. It simplifies to the Stuart-Landau equation~\cite{landau1944problem,stuart1958non}, commonly used to investigate nonlinear oscillator systems near a Hopf bifurcation.
	
	We can further analyze the corrections to this lower harmonic $a_0$ from the interaction between $a_0$ with $a_1$. Keeping $m,n,p\leq M=1$ leads to
	\begin{align}
		\langle x^{2}\dot{x}\rangle_{k=0}^{c}=&-a_{0}^*|a_0|^2+a_{0}^{2}a_{1}^{*}-2a_{1}a_{0}^{*}a_{1}\,.
		\intertext{Similarly, keeping $m,n,p\leq M=2$, introduces additional corrections,}
		\langle x^{2}\dot{x}\rangle_{k=0}^{c}=&a_{0}^{2}a_{1}^{*}-a_{0}\left(a_{0}^{*}{}^{2}-2a_{1}a_{2}^{*}\right)-2a_{1}a_{0}^{*}a_{1}^{*}-a_{2}\left(a_{1}^{*}{}^{2}+2a_{0}^{*}a_{2}^{*}\right)\,,\label{eq:M2}\\
		\intertext{where $a_0$ and $a_2$ are directly coupled, but also $a_0$, where $a_0$, $a_1$ and $a_2$ have a threefold interaction. At increasing orders, nonlinear contributions between all harmonics proliferate, see}
		\langle x^{2}\dot{x}\rangle_{k=0}^{c}=&a_0^2 a^*_1 + a_0 \left(2 (a_1 a^*_2 + a_2 a^*_3 + a_3 a^*_4) - a^*_0{}^2\right) - a_2 a^*_1{}^2 - a_4 a^*_2{}^2 - 2 a_2 a^*_0 a^*_2 - 2 a_3 a^*_1 a^*_2 + a_1^2 a^*_3 -\nonumber\\
		&2 a^*_3 (a_3 a^*_0 + a_4 a^*_1) + a_1 (2 a_2 a^*_4 - 2 a^*_0 a^*_1)\,.
	\end{align}
	The Fourier analysis [cf.~Fig.2(b) in main text] shows $a_k\mapsto 0$ as $k$ increases. Given the prefactors in that the terms above are comparable, the convergence of the ansatz entails that the terms $a_m a_n a_p$ are increasingly weaker as they contain harmonics for larger $m,n,p$, and the result is expected to converge. However, further analysis is needed to understand if such convergence is monotonic. In Fig.~\ref{fig:SI1} we show how, in fact, this is not the case: Increasing the number of harmonics in the ansatz $M$ does not necessarily increase the fidelity of the solution. 
	
	\begin{figure}[t]
		\includegraphics[width=\linewidth]{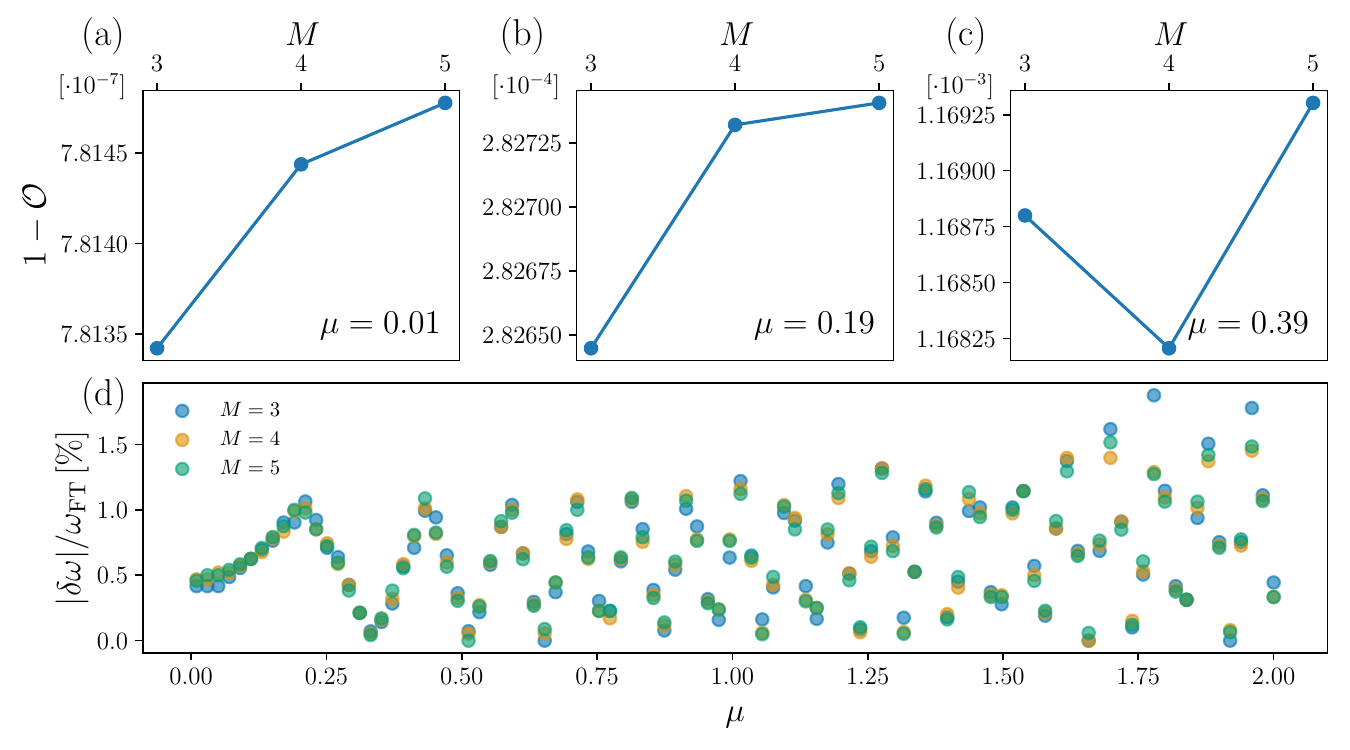}
		\caption{\textit{Fidelity of e-HBM ansatz, further details:} (a)-(c) Overlap between e-HBM and exact time evolution result as a function of the number of harmonics $M$ for varying $\mu$, shown at each panel [cf.~Fig.~2 in the main text]. The overlap is computed by matching frequency peaks in $|\tilde{x}(\omega)|^2$, from the quasi-exact evolution $\textbf{a}_{\text{FT}}$ (cf.~main text) to the nearest frequency estimates obtained by the HBM $\textbf{a}_{\text{HBM}}$, with percentile deviations below $2\%$, shown in (d). } \label{fig:SI1}
	\end{figure}

\end{widetext}

\end{document}